# Optimal control of a single leg hopper by Liouvillian system reduction

Patrick Slade, Siobhan Powell, Michael F. Howland

*Abstract*— The benefits of legged locomotion shown in nature overcome challenges such as obstacles or terrain smoothness typically encountered with wheeled vehicles. This paper evaluates the benefits of using optimal control on a single leg hopper during the entire hopping motion. Basic control without considering physical constraints is implemented through hand-tuned PD controllers following the Raibert control framework. The differential flatness of the first-order equations of motion and the Liouvillian property for the second-order equations for the hopper system are proved, enabling flat outputs for control. A two-point boundary value problem (BVP) is then used to minimize jerk in the flat system to gain implicit smoothness in the output controls. This smoothness ensures that the planned trajectories are feasible, allowing for given waypoints to be reached.

## I. INTRODUCTION

### A. Literature

The agility of animal and human locomotion over rugged terrain has inspired investigation of legged robots. The ability to traverse difficult landscapes could extend robotic applications to environments not accessible to wheeled and tracked robots, impacting applications such as transportation, defense, assistive devices, rescue, and space exploration. Legged locomotion is desirable as a mechanism to transport people, carry heavy loads, and perform exploration on other planets [1], [2], [3], [4].

The leg morphology and method of controlling hopping robots varies widely, including many designs inspired by biology. The pioneering work in single leg hopping separates the control into three parts: (1) forward movement, (2) body attitude, (3) and hopping height, which enables three-dimensional (3D) movement [5], [6]. This work provides a breadth of information on dynamic modeling and control during the stance, flight, and landing phases, typically treated independently. The common leg topologies for single leg hoppers are illustrated in Fig. 1.

Internal motion of the leg has been explored for control of a hopper's body angle during the flight phase [10]. Inertial re-orientation using a moment generated at the center of mass rather than the tail is employed in [11]. They show preliminary work linking sequential jumps together off the ground and obstacles to perform dynamic maneuvers with re-orientation during the flight phase to prepare for the next jump. This design focuses on achieving a better vertical jumping agility and does not include compliance or force-sensing at the foot. Without knowledge of forces applied at

This work is supported by the National Science Foundation Graduate Research Fellowship Program Grant DGE-1656518, the Stanford Graduate Fellowship, and the Stanford School of Engineering Fellowship.

The authors are with the Department of Mechanical Engineering, Stanford University, Stanford, CA 94305 USA (e-mail: patslade@stanford.edu).

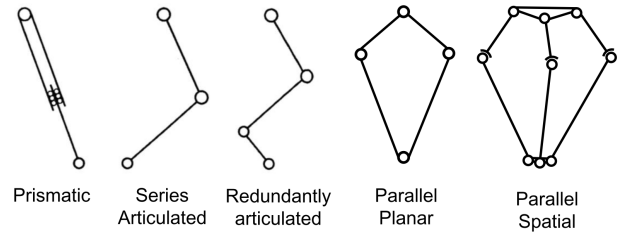

Fig. 1. Common robot leg topologies. These topologies are, from left to right, in the Raibert monopod [6], MIT Cheetah 2 [7], Penn Jerboa [8], Penn/Ghost Robotics Minitaur [8], and the GOAT leg [9]. Figure reprinted with permission of [9]

the foot, the contact with objects and resulting force vector applied when jumping adds uncertainty into the motion of the robot. This increases the difficulty of planning multiple jumps.

The first-order equations of motion (EOM) for a single leg hopper in flight phase were shown to be controllable with differentially flat outputs [12]. That method controls the leg angle with respect to the body, and the rate of change of the leg length. If this were coupled with a BVP during the flight phase to minimize jerk, the rapid change in acceleration from the impact of the leg landing and pushing off would still dominate the trajectory jerk. A more desirable method for optimal control would be minimize jerk across the whole hop.

The Raibert control methods use the second-order EOM to compute a force and torque needed at the foot and hip respectively [6]. Mechanistically, the inputs into the second-order system are more straightforward to implement in a physical system than are the first-order inputs, as they depend directly on measured and controlled quantities rather than estimated parameters.

This differential flatness allows for trajectory generation that can then be optimized in the sense of smoothness. Previous work minimizes the snap, the fourth derivative of position, in quadrotors to pick trajectories that can feasibly reach waypoints [13]. This results in open-loop control which is unable to correct for errors due to noise. Methods such as model predictive control (MPC) add the ability to robustly track these desired open-loop trajectories [14], [15]. Recent techniques improve tracking under uncertainty, when parameters such as friction or mass must be estimated while simultaneously controlling the system [16].

### B. Contribution

Legged hopping locomotion enables traverse of terrain not reachable by wheeled or tracked robots. This hopping move-ment is challenging because the robot acts as an inverted pendulum on a spring, which is a dynamically unstable system. Foot placement during locomotion must be carefully considered to avoid slip and achieve the desired momentum during flight.

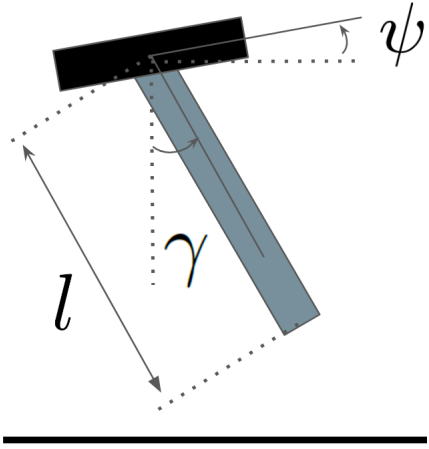

Fig. 2. Depiction of variables used for describing the motion of the single leg hopper

This paper contributes (1) a novel formulation of the second-order EOM for both flight and stance phases as a Liouvillian system with flat output, and (2) solutions to a two-point boundary value problem (BVP) resulting in jerk minimized trajectories for the states and control inputs for optimal hopping.

## II. PROBLEM FORMULATION

Observation of legged locomotion inspires the use of springs and dampers to model the motion of the muscles and tendons. The leg is commonly modeled using a Raibert controller as a spring-damper system. Well known derivations for the spring-loaded-inverted pendulum (SLIP) model describe the dynamical equations for the stance and flight phases of dynamic legged locomotion [17].

### A. SLIP Models

The 1-DOF model of [17] describes the motion of a one-legged hopper as a body point mass, $m$, with a mass-less leg. The leg acts as an ideal linear spring with stiffness, $k$, and no dampening. The motion is constrained to be in the vertical direction, $y$.

$$m\ddot{y} = k(l_0 - y) - mg,$$

where $l_0$ denotes the length of the leg without any spring stretch. The EOM for the *stance* phase in state-space follow as

$$\begin{bmatrix} \dot{y} \\ \ddot{y} \end{bmatrix} = \begin{bmatrix} 0 & 0 \\ -\frac{k}{m} & 0 \end{bmatrix} \begin{bmatrix} y \\ \dot{y} \end{bmatrix} + \begin{bmatrix} 0 \\ \frac{k}{m}l_0 - g \end{bmatrix}.$$

During the *flight* phase, the body mass is only acted on by gravity and follows a falling trajectory where $\ddot{y}$ is equal to the acceleration of gravity $g$.

For the 2-DOF model, the horizontal translation during the flight phase is considered under the same assumptions. The difference between the kinetic, $T$, and potential, $V$, energies in the system allows the Lagrangian dynamics to be defined for the *stance* phase:

$$L = T - V,$$
$$T = \frac{1}{2}m(\dot{l}^2 - l^2\dot{\gamma}^2),$$
$$V = mg(l\sin\gamma) + \frac{1}{2}k(l_0 - l)^2.$$

Substituting $T$ and $V$ into the Lagrange equation and taking the partial derivatives with respect to the state, $r$, gives the result

$$l^2\ddot{\gamma} = gl\cos\gamma.$$

The *flight* phase dynamics remain the same in the y-direction and have no acceleration in the x-direction.

### B. Equations of motion

The EOM for the single leg hopper in this formulation are derived from the Lagrangian in [18]. For our model we will assume $L = 0$, corresponding to the mass centered around the hip joint. These result in the simplified EOM

$$m\ddot{l} - ml\dot{\gamma}^2 + mg\cos(\gamma) = F,$$
$$ml^2\ddot{\gamma} + 2ml\dot{l}\dot{\gamma} - mlg\sin(\gamma) = \tau,$$
$$\ddot{\psi}I_b = \tau,$$

with the inertia of the body represented as $I_b$, the collective force as $F$, and the torque on the system as $\tau$. The resulting Cartesian coordinates of the center of mass and their derivatives are

$$x_{cm} = -l\sin(\gamma),$$
$$y_{cm} = l\cos(\gamma),$$
$$\dot{x}_{cm} = -\dot{l}\sin(\gamma) - l\dot{\gamma}\cos(\gamma),$$
$$\dot{y}_{cm} = \dot{l}\cos(\gamma) - l\dot{\gamma}\sin(\gamma).$$

### C. Raibert hopping controller

An early strategy proposed for hopping robots divides the control into three decoupled 1-DOF controllers which use a state-machine to switch between them [5]. This Raibert controller implements separate PD loops to control the hop height, horizontal hopping speed, and body angle. The hop height is a function of the energy added to the system. Adjusting the length of the hopper in flight controls the compression of the spring on landing. To track the desired horizontal speed, $\dot{x}_d$, the control tries to achieve a desired forward foot position, $x_d$, while in flight

$$x_d = \frac{\dot{x}T_s}{2} + k_{\dot{x}}(\dot{x} - \dot{x}_d).$$

The inverse kinematics are used to calculate the joint angles required for this desired foot position. The desired angle, $\gamma_d$, is then given as

$$\gamma_d = \psi - \sin^{-1}\left(\frac{x_d}{l}\right).$$

The desired hip angle, $\gamma_d$, is tracked with a PD controller with proportional and derivate gains, $k_p$ and $k_v$

$$\tau = -k_p(\gamma - \gamma_d) - k_v(\dot{\gamma}).$$

The desired body angle $\psi_d$ uses a similar PD loop

$$\tau = -k_p(\psi - \psi_d) - k_v(\dot{\psi}).$$

## III. PROPOSED SOLUTION

A two-point BVP based solution to the optimal control problem is implemented within the single legged hopper system following a Raibert framework.

### A. Differential flatness

The property of differential flatness is demonstrated by change of variables which allows the system to be expressed in a non-physical flat output space. All original states and inputs can be explicitly expressed in terms of the flat outputs and a finite number of its derivatives. Differential flatness is useful for generating trajectories or simplifying optimization formulations, but adds difficulty in enforcing physical constraints on the outputs since the original interpretations are obscured by the change of variables defining the flat space.

*1) First-order model:* While the hopper is in the air, the first-order EOM for the simplified monopod hopper are given in [12] as

$$\dot{l} = F,$$
$$\dot{\gamma} = \tau,$$
$$\dot{\psi} = \tau\Big(\frac{m_l(l+l_0)^2}{m_b + m_l(l+l_0)^2}\Big),$$

where $m_l$ and $m_b$ represent the masses of the leg and body. The flat output variables are chosen as $y_1 = \frac{\pi}{2} - \gamma$ and $y_2 = \psi$. This gives the dynamics in terms of the flat outputs as

$$\gamma = \frac{\pi}{2} - y_1,$$
$$\psi = y_2,$$
$$\tau = -\dot{y}_1,$$
$$(l+l_0)^2 = -\frac{m_b \dot{y}_2}{m_l(\dot{y}_2 + \dot{y}_1)},$$
$$F = \frac{m_b(\ddot{y}_1 \dot{y}_2 - \dot{y}_1 \ddot{y}_2)}{2m_l(l+l_0)(\dot{y}_1 + \dot{y}_2)^2}.$$

It can immediately be verified that the choice of $y_1$ and $y_2$ are not differentially related for this system, so we conclude the first-order system is differentially flat [19]. This first-order system would be difficult to implement in hardware since it requires control of velocity rather than acceleration. Thus, a similar flatness result is derived for the second-order system in this paper.

*2) Second-order model:* During *stance*, the second-order EOM for the simplified monopod hopper as derived above are

$$F = m\ddot{l} - ml\dot{\gamma}^2 + mg\cos(\gamma),$$
$$\tau = ml^2\ddot{\gamma} + 2ml\dot{l}\dot{\gamma} - mlg\sin(\gamma),$$
$$\tau = \ddot{\psi}I_b.$$

Equivalently, where $u_2 = \tau/m$ and $u_1 = F/m$,

$$\ddot{l} = l\dot{\gamma}^2 - g\cos(\gamma) + u_1,$$
$$\ddot{\gamma} = \frac{g\sin(\gamma)}{l} - 2\frac{\dot{l}\dot{\gamma}}{l} + \frac{u_2}{l^2},$$
$$\ddot{\psi} = \frac{\tau}{I_b} = \alpha u_2.$$

The flat output variables are selected as $y_1 = l$ and $y_2 = \gamma$. The system variables can be expressed in terms of these flat outputs as

$$l = y_1,$$
$$\gamma = y_2,$$
$$u_1 = \ddot{y}_1 - y_1\dot{y}_2^2 + g\cos(y_2),$$
$$u_2 = y_1^2\ddot{y}_2 + 2y_1\dot{y}_1\dot{y}_2 - gy_1\sin(y_2),$$
$$\psi = \int\int y_1^2\ddot{y}_2 + 2y_1\dot{y}_1\dot{y}_2 - gy_1\sin(y_2)\ dt\ dt.$$

Again it can immediately be verified that the choice of $y_1$ and $y_2$ are not differentially related for this system.

Similarly, in the *air* the EOM are

$$\ddot{l} = 0,$$
$$\ddot{\gamma} = \frac{u_2}{l^2},$$
$$\ddot{\psi} = \frac{\tau}{I_b} = \alpha u_2.$$

Using the same flat output variables, $y_1 = l$ and $y_2 = \gamma$, the dynamics in terms of the flat outputs are given as

$$l = y_1,$$
$$\gamma = y_2,$$
$$u_1 = 0,$$
$$u_2 = y_1^2\ddot{y}_2,$$
$$\psi = \alpha\int\int y_1^2\ddot{y}_2\ dt\ dt.$$

From these equations with the integral relation for $\psi$, it cannot be concluded that the system is differentially flat. The properties instead define a Liouvillian system which can be treated and controlled in the same way [19].

Then the system can be rewritten in terms of $y_1$ and $y_2$ as

$$\frac{d}{dt}\begin{bmatrix}y_1\\y_2\\\dot{y}_1\\\dot{y}_2\end{bmatrix} = \begin{bmatrix}0 & 0 & 1 & 0\\0 & 0 & 0 & 1\\0 & 0 & 0 & 0\\0 & 0 & 0 & 0\end{bmatrix}\begin{bmatrix}y_1\\y_2\\\dot{y}_1\\\dot{y}_2\end{bmatrix} + \begin{bmatrix}0 & 0\\0 & 0\\1 & 0\\0 & 1\end{bmatrix}\begin{bmatrix}v_1\\v_2\end{bmatrix}.$$

The outputs $y_1$, $y_2$, $\dot{y}_1$, and $\dot{y}_2$ are used to back-out $l$, $\psi$, $u_1$, $u_2$, and $\gamma$.

## B. Two-point BVP

Using the second-order Liouvillian control scheme described in Section III-A.2 a control system and objective function are designed to minimize jerk. The jerk terms, $\dddot{y}_1$ and $\dddot{y}_2$ can be expressed in the flat outputs as

$$\frac{d}{dt}\begin{bmatrix} y_1 \\ y_2 \\ \dot{y}_1 \\ \dot{y}_2 \\ \ddot{y}_1 \\ \ddot{y}_2 \end{bmatrix} = \begin{bmatrix} 0 & 0 & 1 & 0 & 0 & 0 \\ 0 & 0 & 0 & 1 & 0 & 0 \\ 0 & 0 & 0 & 0 & 1 & 0 \\ 0 & 0 & 0 & 0 & 0 & 1 \\ 0 & 0 & 0 & 0 & 0 & 0 \\ 0 & 0 & 0 & 0 & 0 & 0 \end{bmatrix} \begin{bmatrix} y_1 \\ y_2 \\ \dot{y}_1 \\ \dot{y}_2 \\ \ddot{y}_1 \\ \ddot{y}_2 \end{bmatrix} + \begin{bmatrix} 0 & 0 \\ 0 & 0 \\ 0 & 0 \\ 0 & 0 \\ 1 & 0 \\ 0 & 1 \end{bmatrix} \begin{bmatrix} w_1 \\ w_2 \end{bmatrix}.$$

The control system can then be written compactly as

$$\frac{d}{dt}\mathbf{y} = A\mathbf{y} + B\mathbf{w},$$

$$\min J = \int_0^\infty \mathbf{w}^T \mathbf{w} \, dt.$$

*1) Hamiltonian:* The minimization of jerk may be formulated as an indirect solution using the Hamiltonian where

$$\mathcal{H} = \mathbf{w}^T \mathbf{w} + P^T A \mathbf{y} + P^T B \mathbf{w}$$

and

$$\dot{x} = \frac{\partial \mathcal{H}}{\partial P} \qquad \dot{P} = -\frac{\partial \mathcal{H}}{\partial x} \qquad 0 = \frac{\partial \mathcal{H}}{\partial w}.$$

The solution of the minimization corresponds to the solution of 12 coupled ordinary differential equations with boundary conditions specified at $t_0$ and $t_f$. The control scheme can be written as

$$\mathbf{w} = \frac{-P^T B}{2}.$$

Since the dynamical system depends on the state (flight or stance) of the hopper, the BVP must be formulated individually for each state. For the stance state, the full 12 equation boundary value problem must be solved. For the air state, the leg length, $l$, is fixed, and the system reduces to a 6 equation boundary value problem. In the air, control $v_1 = 0 \; \forall t$ and $w_1 = 0 \; \forall t$.

*2) Boundary conditions:* In the air state, the solution for $y_2$ is evaluated with the boundary conditions

$$y_2(t_0) = \gamma_0 \qquad y_2(t_f) = \gamma_d,$$

$$\dot{y}_2(t_0) = \dot{y}_2(t_f) = \ddot{y}_2(t_0) = \ddot{y}_2(t_f) = 0.$$

$\gamma_d$ defines the known desired final leg angle for when the hopper returns to the stance phase, $x_f$,

$$x_f = \frac{\dot{x}t_f}{2} + G_\gamma(\dot{x} - \dot{x}_d),$$

$$\gamma_d = \tan^{-1}(\frac{y}{x_f}).$$

Here $G_\gamma$ defines the angle gain, $\dot{x}_d$ the desired hopper lateral speed, and $y$ a known state of the leg. The final time, $t_f$, is fixed and given by equations of projectile motion.

While in the stance phase, boundary conditions must be specified for $y_1$ and $y_2$. Additionally as $t_f$ is free, an additional boundary condition must be specified, giving 13 total. The boundary conditions are:

$$y_1(t_0) = l_0 \qquad y_1(t_f) = l_0$$
$$, y_2(t_0) = \gamma_0 \qquad y_2(t_f) = -\gamma_0 \qquad ,$$
$$\dot{y}_1(t_0) = \dot{l}_{t_0} \qquad \dot{y}_1(t_f) = \dot{l}_{t_f},$$
$$\ddot{y}_1(t_0) = \ddot{y}_1(t_f) = 0,$$
$$\dot{y}_2(t_0) = \dot{y}_2(t_f) = \ddot{y}_2(t_0) = \ddot{y}_2(t_f) = 0,$$
$$\mathbf{w}^T(t_f)\mathbf{w}(t_f) + P^T(t_f)\left[A\mathbf{y}(t_f) + B\mathbf{w}(t_f)\right] = 0.$$

The choice of the boundary conditions for $\dot{y}_1$ ($\dot{l}_{t_0}$ and $\dot{l}_{t_f}$) represents a tuning of the system. Due to the nature of switching dynamical systems in the transition from flight to stance or vice versa, the system is sensitive to the choice of the sign of $\dot{y}_1$; namely, $\dot{y}_1$ must be positive during takeoff and negative during landing. In practice it was found that the system was not very sensitive to the choice of magnitudes as long as the correct sign was specified (not shown).

## IV. SIMULATION

In order to compare the control methods for the monopod hopper, simulations were performed with the model of the single leg hopper presented in Fig. 2. A desired hopping height and horizontal speed were given as reference values. A fourth-order Runge-Kutta dynamics update was implemented in Matlab with small measurement and process noise.

### A. PD

The state trajectories for the hand-tuned PD system are shown in Fig. 3. The change in leg length appears to be quite smooth, but the leg angle has difficulty tracking the desired angle during the flight phase. The hip torque applied during flight phase in Fig. 4 highlights the sharp control changes which can cause jerk in the system due to rapid changes in acceleration. A second-order central differencing equation was used to approximate jerk for the leg angle, as displayed in and Fig. 5. It is important to note that the magnitude of these jerk values is on the order of $10^7$, indicating difficulty in smoothing over this erratic change in the system when moving between waypoints. In practice, these state trajectory functions are non-differentiable when using PD, which leads to large magnitudes of jerk after numerical differencing.

### B. Two-point BVP

The two-point boundary value problem was solved using a built-in Matlab function $bvp4c$ which is a robust adaptive-mesh three-stage Lobatto IIIa direct collocation solver. The $t_f$ free condition was simulated in Matlab using the BVP standard form manipulation of [20]. Typically, the solutions to BVPs are not guaranteed to converge for all initialization conditions, and thus there is some tuning required for solution validity.

Within the overall hopper simulation, the two-point BVP was the most expensive execution, pausing the simulation on the order of 1 second when running in "real time." In order to reduce the computational tasks, we limited the trajectory computation to once per hopper state change (i.e. the BVP is solved when the hopper changes from stance to flight or flight to stance). This choice decreases the cost significantly but makes the hopper more susceptible to higher noise levels, occasionally causing failure. The speed of execution can be considerably improved through methods such as (but not limited to) parallel implementation, lower order methods such as the shooting method, and more refined initialization. In practice if computational costs are prohibitive, a suite of trajectories for many boundary conditions may be computed a priori offline and executed in real time.

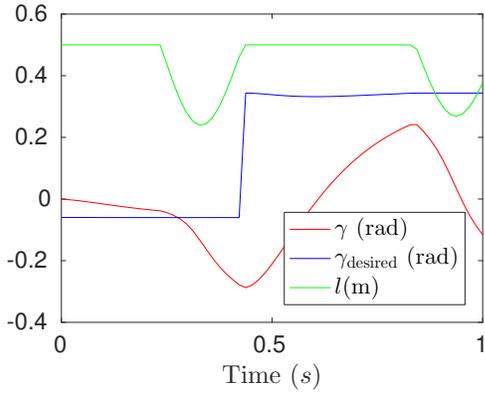

Fig. 3. The state variables for the hand-tuned PD controller for two hops

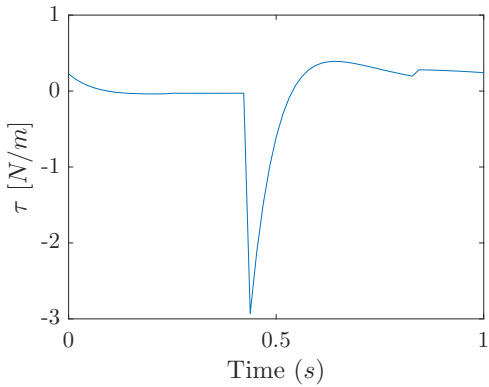

Fig. 4. Hip torque for the hand-tuned PD controller during a typical stance phase

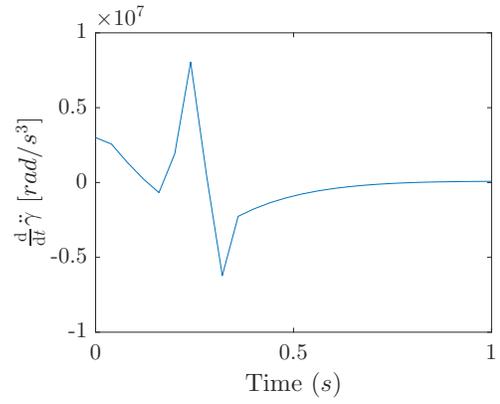

Fig. 5. Jerk in the leg angle for the hand-tuned PD controller during a typical stance phase

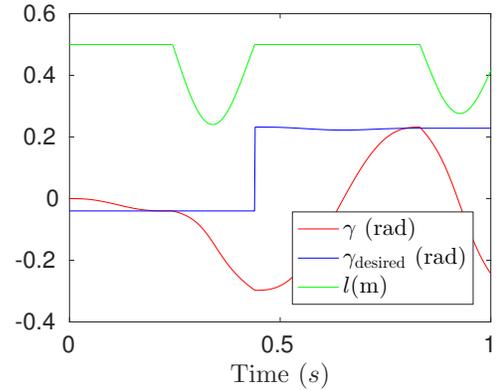

Fig. 6. The state variables for the jerk minimized boundary value problem for two hops

The state trajectories in Fig. 6 highlight a smooth leg length profile and reach the desired leg angle exactly. The PD controller is not able to track the desired $\gamma_f$ in flight. The torque applied to the hip in Fig. 7 shows a smooth profile that minimizes jerk for the leg angle in Fig. 8. The jerk is decreased by many orders of magnitude in the optimal control framework. This minimized jerk improves the ability of the hopper to successfully locomote between desired waypoints along a planned route. In practice, the low jerk and smooth torque profiles are significantly better for physical motors and systems, especially with high environmental noise. Additionally, the magnitudes of applied control are significantly reduced for the jerk minimized system even though it was not directly incorporated into the cost function. Since the model has been shown to be Louivillian, it is relatively simple to incorporate most sophisticated cost penalties beyond jerk minimization such as energy or control effort.

With no noise included in the simulation, the control provided by the BVP in the flat space satisfies the boundary conditions to numerical precision, ensuring that the trajectory is feasible. When noise is incorporated, the state and control trajectories become uncertain and feasibility is not guaranteed with the BVP solution.

## V. CONCLUSION

This paper proves that the second-order EOM for a single legged hopper define a Liouvillian system. The states and inputs are explicitly expressed in terms of flat outputs and a finite number of their derivatives, allowing state and control trajectories to be found. To ensure that trajectories are feasible when planning between waypoints, the jerk in the flat system is minimized to add implicit smoothness that acts similar to bounds on the control output. The reduced order Liouvillian system allows for a variety of optimal control methods to be more easily implemented and is not limited to jerk minimization.

This work will be extended by augmenting the open loop control with MPC. Tests to determine noise rejection and stability with this formulation should be explored.

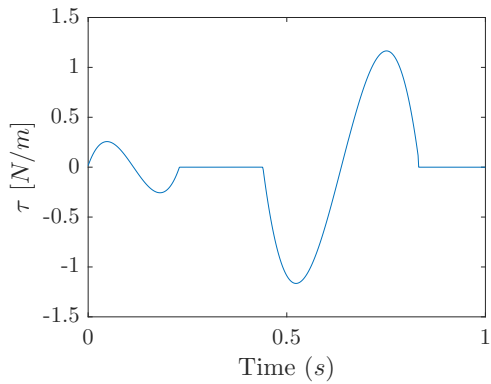

Fig. 7. Hip torque for the jerk minimized boundary value problem during a typical stance phase

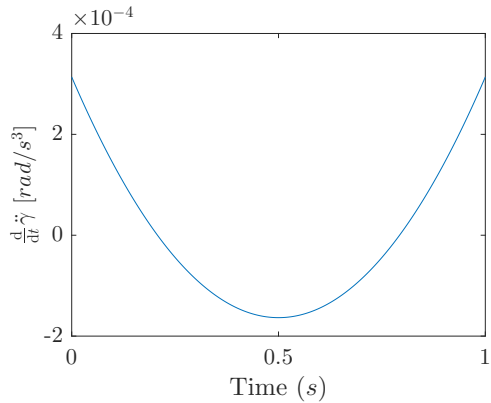

Fig. 8. Jerk in the leg angle for the jerk minimized boundary value problem during a typical stance phase